\documentclass[english,pra,aps,preprint,showpacs]{revtex4-1}
\usepackage[T1]{fontenc}
\usepackage[latin9]{inputenc}
\usepackage[active]{srcltx}
\usepackage{amsmath}
\usepackage{amssymb}
\usepackage{esint}

\makeatletter

\@ifundefined{textcolor}{}
{%
 \definecolor{BLACK}{gray}{0}
 \definecolor{WHITE}{gray}{1}
 \definecolor{RED}{rgb}{1,0,0}
 \definecolor{GREEN}{rgb}{0,1,0}
 \definecolor{BLUE}{rgb}{0,0,1}
 \definecolor{CYAN}{cmyk}{1,0,0,0}
 \definecolor{MAGENTA}{cmyk}{0,1,0,0}
 \definecolor{YELLOW}{cmyk}{0,0,1,0}
}

\usepackage{epsf}
\usepackage{epsfig}
\usepackage{graphicx}

\newcommand{\ket}[1]{|#1\rangle}

\newcommand{\outprod}[2]{|#1\rangle\langle#2|}
\newcommand{\half}{\mbox{$\textstyle \frac{1}{2}$}}

\newcommand{\Rtil}{\tilde{\rho}}

\makeatother

\begin{document}

\title{Post Markovian Dynamics of Quantum Correlations: Entanglement vs.
Discord}

\author{Hamidreza Mohammadi}

\thanks{hr.mohammadi@sci.ui.ac.ir}

\affiliation{Department of Physics, University of Isfahan, Isfahan, Iran}

\affiliation{Quantum Optics Group, University of Isfahan, Isfahan, Iran}

\begin{abstract}
\noindent Dynamics of an open two-qubit system is investigated in the post-Markovian regime, where the environments have a short-term memory. Each qubit is coupled to separate environment which is held in its own temperature. The inter-qubit interaction is modeled by XY-Heisenberg model in the presence of spin-orbit interaction and inhomogeneous magnetic field. The dynamical behavior of entanglement and discord has been considered. The results show that, quantum discord is more robust than quantum entanglement, during the evolution. Also the asymmetric feature of quantum discord can be monitored by introducing the asymmetries due to inhomogeneity of magnetic field and temperature difference between the reservoirs. By employing proper parameters of the model, it is possible to maintain non-vanishing quantum correlation at high degree of temperature. The results can provide a useful recipe for studying of dynamical behavior of two-qubit systems such as trapped spin-electrons in coupled quantum dots. 
\end{abstract}


\maketitle

\section{INTRODUCTION}
The weirdness of quantum mechanics lies on the concept of quantum correlation which is originated from the superposition principle.  There are two important aspects of quantum correlation: \textit{quantum entanglement} \cite{EPRPRA1935,SchrodingerNat1935}which is defined within the entanglement-separability paradigm and \textit{quantum discord} \cite{ZurekPRL2001,HendersonJPA2001}, defined from an information-theoretic perspective.
The quantitative and qualitative evaluation of such correlations is central task in conceptual studies of conceptual quantum mechanics, and it also has crucial significance in operative quantum information theory. A mixed state $\rho$ of a bipartite system is an entangled state if it not separable i.e. it can not prepared by Local Operation and Classical Communication (LOCC) tasks. There are many measures which evaluate the amount of entanglement of a quantum state.   The entanglement of formation is one of these measures, which is enumerates the resources which is needed to create a given entangled state. For the case of a two-qubit bipartite system the formula of the entanglement of formation can be expressed as a smooth function of \textit{the concurrence} and hence the concurrence can be take as a measure of entanglement in its own right\cite{WoottersPRL1997}. The concurrence of the state $\rho_{AB}$ can be obtained
explicitly as:
\begin{equation} 
C(\rho_{AB})=\max\{0,2\lambda_{max}-\sum_{i=1}^{4}\,\lambda_{i}\},\label{concurrence}
\end{equation}
where $\lambda_{i}$s are roots of the eigenvalues of the non-Hermitian
matrix $R=\sqrt{\sqrt{\rho_{AB}}\Rtil_{AB}\sqrt{\rho_{AB}}}$, and $\tilde{\rho}_{AB}$ is defined
by $\Rtil_{AB}:=(\sigma^{y}\otimes\sigma^{y})\rho_{AB}^{*}(\sigma^{y}\otimes\sigma^{y})$, here $\sigma^y$ is Pauli y-matrix \cite{WoottersPRL1997}. 

Until some time ago, entanglement was considered as the only type
of quantum correlation could be find in a composed quantum state. However,
it has been discovered that some multi-partite separable states could speedup quantum computation algorithms \cite{KnillPRL1998} i.e. they possess some quantum features. Therefore, entanglement is not the only aspect of quantum correlation. Datta {\it et. al.} \cite{DattaPRL2008} have showen that the resource of this speedup is  another important type of quantum correlations, named by Quantum Discord (QD). Quantum discord first introduced by Zurek {\it et. al.}   \cite{ZurekPRL2001} and Henderson {\it et. al.} \cite{HendersonJPA2001}, independently in year 2001. The definition of QD lies on the difference between two classically equivalent definitions of mutual information in the quantum mechanics language. In mathematical sense the quantum  discord could be obtained by eliminating the classical correlation from the total correlation measured by quantum mutual information. The classical correlation between the parts of a bipartite system can be obtained by use of the measurement-base conditional density operator. Hence we can write the discord with respect to the $B$ subsystem (right discord) as:
\begin{equation}
D_{B}(\rho_{AB})=I(\rho_{AB})-CC_{B}(\rho_{AB}).\label{discord}
\end{equation}
Where $I(\rho_{AB})=S(\rho_{A})+S(\rho_{B})-S(\rho_{AB})$ is the mutual information and $CC_{B}(\rho_{AB})=\underset{\{\Pi_{k}^{(B)}\}}{\sup}\{S(\rho_{A})-S(\rho_{AB}|\{\Pi_{k}^{(B)}\})\}$ is the classical part of correlation. Here $\rho_{A(B)}$ and $\rho_{AB}$ refer to the reduced density matrix of subsystem $A(B)$ and the density matrix of the system as the whole and $S(\rho)=-Tr(\rho\,\log_{2}\rho)$ is Von Neumann entropy. The maximization in the definition of classical correlation is taken over the set of generalized measurements (POVMs) $\{\Pi_{k}^{(B)}\}$, and $S(\rho_{AB}|\{\Pi_{k}^{(B)}\})=\sum_{k=0}^{1}p_{k}S(\rho_{k})$ is
the conditional entropy of subsystem $A$, with $\rho_{k}=Tr_{B}((I_{A}\otimes\Pi_{k}^{(B)})\,\:\rho_{AB}\,\:(I_{A}\otimes\Pi_{k}^{(B)}))/p_{k}$ and $p_{k}=Tr(\rho_{AB}\:(I_{A}\otimes\Pi_{k}^{(B)}))$. However, one can swap the role of the subsystems $A$ and $B$ to
obtain discord with respect to $A$ subsystem (left discord), i.e. $D_{A}(\rho_{AB})$.

Decoherence is the main obstacle to preserving the superposition and hence quantum correlation in real quantum systems. Undesired leakage of the coherence of the system to the environment, due to
unavoidable interaction between the quantum systems and their environment,
leading to decoherence \cite{Schlosshauerbook2007}. Thermal decoherence plays a significant role to destroying the useful quantum correlation between the parts of the quantum systems. Although investigating the decohernce procedure in thermal equilibrium is useful but real systems are not in equilibrium \cite{VedralJP2009} and hence the dynamical behavior of the systems under non-equilibrium condition has to be elucidated.
Furthermore, the formal analysis of an open quantum dynamics is considered in Markovian framework i.e. by assuming the weak system-environment coupling and the forgetful nature of the environmental system. Despite of its wide applicability, it should be kept in mind that Markovianity is only an approximation and the real physical systems may not fulfill these conditions. This imposes one to address the question of quantum feature survival in noisy as well as non-equilibrium conditions in the non-Markovian regime.

In propose to realize such systems we consider the non-equilibrium dynamics
of a system including two coupled qubits in contact with different
thermal baths. This is a system which is interesting both from theoretical
and empirical point of view. Recent progresses in nano-technology provide the possibility of fabrication and manipulation of confined spins in nano-scale devices. Among these devices, semiconductor quantum dots becomes a useful device for manipulating, transferring and saving the quantum information. For example, data transferring between nuclear spins and electronic spins confined in a semiconductor quantum dot has been considered \cite{ReinaPRB2000}. These nuclear and the electronic are embedded in a solid state environment with huge degrees' of freedom (bath). Because they interact with their bath in completely different manner, they experience different effects from environment. Nowadays, with the aid of NMR and quantum optical techniques, it is possible to change the temperature of the nuclear spins without affecting the electron spins  temperature \cite{StepankoPRL2006}. A system consists of two spin-electron confined in two coupled quantum dots\cite{LossPRA1998, DiVincenzoPRA1995}, is another example. In this system qubit is represented by a single spin-electron confined in each quantum dot. These qubits can be initialized, manipulated, and read out by extremely sensitive devices. In comparison with quantum optical and NMR systems, such systems are more scalable and more robust to the environmental affects. Each quantum dot could be coupled to different source and drain electrodes, during the fabrication process, and hence feels a different environment \cite{LegelPRB2007}.

In this paper, the non-equilibrium dynamics of an open quantum system is investigated. The system to be considered includes a two-qubit system interacting with surrounding environment. The inter-qubit separation is supposed large enough such that each qubit is embedded in a separate environment. The environments are modeled by thermal reservoirs (bosonic bathes) which are assumed to be in thermodynamic equilibrium in their own temperature $\beta_{i}=\frac{1}{k_{B}T_{i}}$. Furthermore each qubit is realized by the spin of an electron confined in a quantum dots, so  due to weak lateral confinement, electrons can tunnel from one dot to the other and spin-spin and spin-orbit interactions between the two qubits exist. Also, it is assumed that an external magnetic field is applied to each quantum dot. Thus the inter-qubit interaction could be modeled by anisotropic XY Heisenberg system in the presence of the inhomogeneous magnetic field, equipped by spin-orbit interaction in the form of the Dzyaloshinski-Moriya (DM) interaction \cite{KavokinPRB2001, DzyaloshinskiJPCS1958, MoriyaPRL1960}.
In the following, the influence of the parameters of the system (i.e.
magnetic field (B), inhomogeneity of magnetic field (b), partial anisotropy($\chi$),
mean coupling (J) and the spin-orbit interaction parameter (D)) and
environmental parameters (i.e. temperatures $T_{1}$ and $T_{2}$
or equally $T_{M}$ and $\Delta T$, and the couplings strength $\gamma_{1}$
and $\gamma_{2}$) on the amount of entanglement and discord of the
system is studied. The results show that the dynamics of quantum correlations depends on the geometry of connection, especially the geometry of connection can bold asymmetric property of quantum discord. Also the results show that, for an exponentially damping
memory kernel, there is a steady state in asymptotically large time limit. The amount of both asymptotic entanglement and asymptotic discord decreases as the temperature increases
but for asymptotic discord sudden death does not occur;  asymptotic discord descends
exponentially with temperature, while the entanglement suddenly vanishes
above a critical temperature, $T_{M}^{cr.}$. The results also reveal
that, the size of $T_{M}^{cr.}$ (temperature over which the quantum
entanglement cease to exist) and the amount of both entanglement and
discord can be improved by adjusting the value of the spin-orbit interaction
parameter $D$. This parameter can be manipulated by adjusting the height of the barrier between two quantum dots.

The paper is organized as follows. In Sec. II, we introduce the Hamiltonian
of the whole system-reservoir and then write the post-Markovian master
equation governed on the system by tracing out the reservoirs' degrees
of freedom. Ultimately, for X-shaped initial states, the density matrix
of the system at a later time is derived exactly. The effects of initial
conditions and system parameters on the dynamics of entanglement and
entanglement of asymptotic state of the system are presented in Sec.
III. Finally in Sec. IV a discussion concludes the paper.

\section{THE MODEL AND HAMILTONIAN}

A bipartite quantum system coupled to two
reservoirs is described by the following Hamiltonian: 
\begin{eqnarray}
\hat{H}=\hat{H}_{S}+\hat{H}_{B1}+\hat{H}_{B2}+\hat{H}_{SB1}+\hat{H}_{SB2},\label{total Hamiltonian}
\end{eqnarray}
where $\hat{H}_{S}$ is the Hamiltonian of the system, $\hat{H}_{Bj}$
is the Hamiltonian of the jth reservoir $(j=1, 2)$ and $\hat{H}_{SBj}$ denotes
the interaction Hamiltonian of the system and jth reservoir. According to previous section, the system under consideration consists of two interacting spin electrons confined in two coupled quantum dots. Thus inter-qubit interaction includes spin-spin interaction and spin-orbit interaction (due to orbital motion of electrons). Such system could be described by
a two-qubit anisotropic Heisenberg XY-model in the presence of an
inhomogeneous magnetic field equipped by spin-orbit interaction with
the following Hamiltonian (see \cite{HamidPRA2008} and references therein):
\begin{equation}
\hat{H}_{S}={\textstyle \frac{1}{2}}(J_{x}\,\sigma_{1}^{x}\sigma_{2}^{x}\,+J_{y}\,\sigma_{1}^{y}\sigma_{2}^{y}+\textbf{B}_{1}\cdot\boldsymbol{\sigma}_{1}+\textbf{B}_{2}\cdot\boldsymbol{\sigma}_{2}+\boldsymbol{D}.(\boldsymbol{\sigma}_{1}\times\boldsymbol{\sigma}_{2})),\label{system Hamiltonian}
\end{equation}
 where $\boldsymbol{\sigma}_{j}=(\sigma_{j}^{x},\sigma_{j}^{y},\sigma_{j}^{z})$
is the vector of Pauli matrices, $\textbf{B}_{j}\,(j=1,2)$ is the
magnetic field on site j, $J_{\mu}s\,(\mu=x,y)$ are the real coupling
coefficients (the interaction is anti-ferromagnetic (AFM) for $J_{\mu}>0$
and ferromagnetic (FM) for $J_{\mu}<0$) and $\boldsymbol{D}$ is
Dzyaloshinski-Moriya term of spin-orbit interaction. Reparametrizing the above Hamiltonian with $\textbf{B}_{j}=B_{j}\,\hat{\boldsymbol{z}}$ such that $B_{1}=B+b$ and $B_{2}=B-b$, where b is magnetic field
inhomogeneity, and $J:=\frac{J_{x}+J_{y}}{2}$, as the mean coupling coefficient
in the XY-plane, $\chi:=\frac{J_{x}-J_{y}}{J_{x}+J_{y}}$, as partial anisotropy, $-1\leq\chi\leq1$,
and with the assumption $\boldsymbol{D}=J\, D\hat{\boldsymbol{\, z}}$ we have:
\begin{eqnarray}
\hat{H}_{S} & = & J\chi(\sigma_{1}^{+}\sigma_{2}^{+}+\sigma_{1}^{-}\sigma_{2}^{-})+J(1+i\, D)\,\sigma_{1}^{+}\sigma_{2}^{-}+J(1-i\, D)\,\sigma_{1}^{-}\sigma_{2}^{+}\nonumber \\
 & + & (\frac{{B+b}}{2})\sigma_{1}^{z}+(\frac{{B-b}}{2})\sigma_{2}^{z},\label{system Hamiltonian1}
\end{eqnarray}
where  $\sigma^{\pm}=\frac{1}{2}(\sigma^{x}\pm i\sigma^{y})$, denote the lowering
and raising operators. The spectrum of $\hat {H}_{S}$, in the standard basis
$\{\ket{00},\ket{01},\ket{10},\ket{11}\}$, is easily obtained as
\begin{eqnarray}
\begin{array}{l}
\ket{\varepsilon_{1,2}}=\ket{\Psi^{\pm}}= \sin\theta_{\pm} e^{ i \phi}\ket{01}+\cos\theta_{\pm}\ket{10}\,,\,\;\;\;\;\varepsilon_{1,2}=\pm\xi\,,\\
\\
\ket{\varepsilon_{3,4}}=\ket{\Sigma^{\pm}}=\sin\theta'_{\pm}\ket{00}+\cos\theta'_{\pm}\ket{11}\,,\,\;\;\;\;\;\;\;\varepsilon_{3,4}=\pm\eta\,.
\end{array}\label{spectrum}
\end{eqnarray}
 Where $\tan\theta_{\pm}=\sqrt{\pm(\frac{\xi\pm b}{\xi \mp b})}$, $\tan\phi=D$ and $\tan\theta'_{\pm}=\sqrt{\pm(\frac{\eta\pm B}{\eta \mp B})}$ with $\xi=(b^{2}+J^{2}(1+D^{2}))^{1/2}$ and $\eta=(B^{2}+(J \chi)^2)^{1/2}$. The Hamiltonian of the reservoir coupled to jth spin are given by
\begin{eqnarray}
\hat{H}_{Bj}=\sum_{n}\omega_{n}\hat{a}_{nj}^{\dag}\hat{a}_{nj}\,,\label{Bath Hamiltonian}
\end{eqnarray}
where $\hat{a}_{nj}^{\dag}$  and $\hat{a}_{nj}$ are the creation and the annihilation 
operators of the jth bath mode, respectively. In the full dissipative regime and
in the absence of dephasing processes the interaction between the
system and the jth bath is governed by the following Hamiltonian\cite{HamidEPJD2010}:
\begin{eqnarray}
\hat{H}_{SBj}=(\sigma_{j}^{+}+\sigma_{j}^{-})(\sum_{n}g_{n}^{(j)}\hat{a}_{n,\, j}+g_{n}^{(j)*}\hat{a}_{n,\, j}^{\dag})=\sum_{\mu}(\hat{\Lambda}{}_{j,\,\mu}^{+}+\hat{\Lambda}{}_{j,\,\mu}^{-})(\hat{G}_{j,\,\mu}+\hat{G}_{j,\,\mu}^{\dagger}),\label{Interaction Hamiltonian}
\end{eqnarray}
The system operators $\hat{\Lambda}_{j,\,\mu}^{\pm}$ are chosen to satisfy
$[\hat{H}_{S},\hat{\Lambda}_{j,\,\mu}^{\pm}]=\pm\omega_{j,\,\mu}\hat{\Lambda}_{j,\,\mu}^{\pm}$,
and the $\textit{\^{G}}_{j,\,\mu}$'s are the random operators of
reservoirs and act on the bath degrees of freedom. The Greek letter indexes are related to the transitions between the internal levels of the system induced by the bath. The irreversibility hypothesis implies that the evolution of the system does not influence the states of the reservoirs and the state of whole system+reservoirs is describing by, $\hat{\sigma}(t)=\hat{\rho}(t)\hat{\rho}_{B1}(0)\hat{\rho}_{B2}(0)$, where $\hat{\rho}(t)$ is the reduced density matrix describing the system and each bath is supposed to be in their thermal state at temperature $ T_j=\frac{1}{\beta_j}$, i.e.  $\hat{\rho}_{Bj}=\emph{e}^{-\beta_{j}\hat{H}_{Bj}}/Z$, where $Z=Tr(\emph{e}^{-\beta_{j}\hat{H}_{Bj}})$ is the partition function of the jth bath. Dynamics of the reduced density matrix of system in the Post-Markov approximation is describing with the following master equation\cite{ShabaniPRA2005,DieterichNat2015, CampbellPRA2012, BudiniPRE2014, Sinaysky}:
\begin{eqnarray}
\frac{d\hat{\rho}}{dt}=-i[\hat{H}_{S},\hat{\rho}]+\mathcal{L}\int_{0}^{t}dt'[k(t')\exp(\mathcal{L}\, t')\hat{\rho}(t-t')],\label{master equation}
\end{eqnarray}
 where $\mathcal{L}=\mathcal{L}_{1}+\mathcal{L}_{2}$ and $\mathcal{L}_{j}(\hat{\rho})\,(j=1,2)$
is \textit{dissipator} or \textit{Lindbladian} given by  
\begin{eqnarray}
\mathcal{L}_{j}(\hat{\rho})\equiv\sum_{\mu,\,\nu}J_{\mu,\,\nu}^{(j)}(\omega_{j,\,\nu})\{[\hat{\Lambda}_{j,\,\mu}^{+},[\hat{\Lambda}_{j,\,\nu}^{-},\hat{\rho}]]-(1-\emph{e}^{\beta_{j}\omega_{j,\,\nu}})[\hat{\Lambda}_{j,\,\mu}^{+},\hat{\Lambda}_{j,\,\nu}^{-}\hat{\rho}]\}.\label{dissipiators 1}
\end{eqnarray}
Here $J_{\mu,\nu}^{(j)}(\omega_{j,\nu})$ is the spectral density
of the jth reservoir given by: 
\begin{eqnarray}
J_{\mu,\,\nu}^{(j)}(\omega_{j,\,\nu})=\int_{0}^{\infty}d\tau\emph{e}^{i\omega_{j,\,\nu}\tau}  \langle\textit{\={G}}_{j,\,\mu}(\tau)\,\,\textit{\^G}_{j,\,\nu}\rangle_{\rho_{Bj}},\label{spectral density}
\end{eqnarray}
with $\textit{\={G}}_{j,\,\nu}(\tau)=\emph{e}^{-iH_{Bj}\tau}\textit{\^G}_{j,\,\mu}^{\,\,\,\dag}\emph{e}^{iH_{Bj}\tau}$.
For the bosonic thermal bath modeled by an infinite set of harmonic oscillators, the Weisskpof-Wignner-like approximation implies that: $J^{(j)}(\omega_{\mu})=\gamma_{j}(\omega_{\mu})n_{j}(\omega_{\mu})$  with the property of $J^{(j)}(-\omega_{\mu})=\emph{e}^{\beta_{j}\omega_{\mu}}J^{(j)}(\omega_{\mu})$, where $n_{j}(\omega_{\mu})=(\emph{e}^{\beta_{j}\omega_{\mu}}-1)^{-1}$ is the thermal mean value of the number of excitation in the $j$th reservoir at frequency $\omega_{\mu}$ and $\gamma_{j}(\omega_{\mu})$ is the coupling coefficient of system and the $j$th reservoir. Thus, we can write:
\begin{eqnarray}
\mathcal{L}_{j}(\hat{\rho}) & = & \sum_{\mu=1}^{4}J^{(j)}(-\omega_{\mu})(2\hat{\Lambda}_{j,\,\mu}^{+}\hat{\rho}\hat{\Lambda}_{j,\,\mu}^{-}-\{\hat{\rho},\hat{\Lambda}_{j,\,\mu}^{-}\hat{\Lambda}_{j,\,\mu}^{+}\}_{+}))\nonumber \\
 & + & \sum_{\mu=1}^{4}J^{(j)}(\omega_{\mu})(2\hat{\Lambda}_{j,\,\mu}^{-}\hat{\rho}\hat{\Lambda}_{j,\,\mu}^{+}-\{\hat{\rho},\hat{\Lambda}_{j,\,\mu}^{+}\hat{\Lambda}_{j,\,\mu}^{-}\}_{+})),
\end{eqnarray}
with the transition frequencies
\begin{eqnarray}
\omega_{1}=\xi-\eta=-\omega_{4},\,\,\,\,\,\,\,\,\,\,\,\,\,\,\,\ \omega_{2}=\xi+\eta=-\omega_{3},
\end{eqnarray}
 and the transition operators 
\begin{eqnarray}
\hat{\Lambda}_{j,\,1}^{+} & = & c_{j,\,1}\outprod{\Psi^{+}}{\Sigma^{+}},\,\,\,\,\,\,\,\,\,\,\,\,\,\,\,\,\,\,\,\,\,\,\ \hat{\Lambda}_{j,\,2}^{+} = c_{j,\,2}\outprod{\Psi^{+}}{\Sigma^{-}}, \nonumber \\
\hat{\Lambda}_{j,\,3}^{+} & = & c_{j,\,3}\outprod{\Psi^{-}}{\Sigma^{+}},\,\,\,\,\,\,\,\,\,\,\,\,\,\,\,\,\,\,\,\,\,\,\ \hat{\Lambda}_{j,\,4}^{+}=c_{j,\,4}\outprod{\Psi^{-}}{\Sigma^{-}},\label{operators}
\end{eqnarray}
where 
\begin{eqnarray}
\mid c_{j,\,1}\mid^{2} & = & \mid c_{j,\,4}\mid^{2}=\frac{1}{2\xi\eta}(\xi\eta+J^{2}\chi+(-1)^{j} B b),\nonumber \\
\mid c_{j,\,2}\mid^{2} & = & \mid c_{j,\,3}\mid^{2}=\frac{1}{2\xi\eta}(\xi\eta-J^{2}\chi-(-1)^{j} B b),
\end{eqnarray}
and $\hat{\Lambda}_{j,\,\mu}^{-}=(\hat{\Lambda}_{j,\,\mu}^{+})^\dagger$. Note that, the transition operators $\hat{\Lambda}_{j,\,\mu}^{\pm}$ defined in Eq. (\ref{operators}) just describe the energy exchange between
the system and environment (dissipative coupling), including both
excitation and de-excitation of the qubits. The absence of the transitions
$\Sigma^{+}\leftrightarrow\Sigma^{-}$ and $\Psi^{+}\leftrightarrow\Psi^{-}$
stems in the omittance of the dephasing processes in the system-bath
interaction Hamiltonian. In addition in the rest of the paper, the non-dispersive coupling coefficient is considered i.e.  $\gamma_{j}(\omega_{\mu})=\gamma_{j}$. 

An analytical solution of master equation (\ref{master equation}) can
be obtained by solving the eigenvalue equation $\mathcal{L}\rho=\lambda\rho$.
In this order, the Lindblad superoperator diagonalized with the aid
of its Jordan decomposition form $J$ with $\mathcal{L}=SJS^{-1}$. Fortunately,
the master integro-differential equation (\ref{master equation})
has an important property, when the spectrum of $\hat{H}_{s}$ (see
eq.(\ref{spectrum})) is non-degenerate, the equations for diagonal elements of density matrix decouple from non-diagonal ones \cite{Breuerbook2002}. Thus for the case $\xi\neq \eta$, where the spectrum (\ref{spectrum}) is not degenerate, we can consider them separately. The Lindbladian
for diagonal terms can be written as a time independent $4\times4$
matrix in the energy basis $\{\ket{\varepsilon_{i}}\}_{i=1}^{4}$:
\begin{eqnarray}
\mathcal{L}^{diag}=\left({\begin{array}{cccc}
-(X_{1}^{-}+Y_{2}^{-}) & 0 & X_{1}^{+} & Y_{2}^{+}\\
0 & -(X_{1}^{+}+Y_{2}^{+}) & Y_{2}^{-} & X_{1}^{-}\\
X_{1}^{-} & Y_{2}^{+} & -(X_{1}^{+}+Y_{2}^{-}) & 0\\
Y_{2}^{-} & X_{1}^{+} & 0 & -(X_{1}^{-}+Y_{2}^{+})
\end{array}}\right),\label{B explicit form}
\end{eqnarray}
 where
\begin{eqnarray}
X_{\mu}^{\pm}=2\sum_{j=1,2}J^{(j)}(\mp\omega_{\mu})\mid a_{j,\,1}\mid^{2},\nonumber \\
Y_{\mu}^{\pm}=2\sum_{j=1,2}J^{(j)}(\mp\omega_{\mu})\mid a_{j,\,2}\mid^{2}.
\end{eqnarray}
The Jordan form of this matrix can be obtained easily as:
\begin{equation}
\mathcal{L}^{diag}=SJ^{(d)}S^{-1},\nonumber
\end{equation}
with
\begin{eqnarray*}
S=\left({\begin{array}{cccc}
\frac{Y_{2}^{+}}{Y_{2}^{-}} & \frac{Y_{2}^{+}}{Y_{2}^{-}} & -1 & -1\\
\frac{X_{1}^{-}}{X_{1}^{+}} & -1 & \frac{X_{1}^{-}}{X_{1}^{+}} & -1\\
\frac{X_{1}^{-}}{X_{1}^{+}}\frac{Y_{2}^{+}}{Y_{2}^{-}} & -\frac{Y_{2}^{+}}{Y_{2}^{-}} & -\frac{X_{1}^{-}}{X_{1}^{+}} & 1\\
1 & 1 & 1 & 1
\end{array}}\right),
\end{eqnarray*}
and
\begin{eqnarray*}
J^{(d)}=diag[J_{11}^{(d)}=0,J_{22}^{(d)}=-X_{1},J_{33}^{(d)}=-Y_{2},J_{44}^{(d)}=-(X_{1}+Y_{2})].
\end{eqnarray*}
Knowing the eigenvalues of Linbladian superoperator, $\lambda_{i}^{(d)}=J_{ii}^{d}$
and the memory kernel $k(t)$, the function $\xi_{i}^{(d)}(t)=\xi(\lambda_{i}^{(d)},t)=Lap^{-1}[\frac{1}{s-\lambda_{i}^{(d)}k(s-\lambda_{i}^{(d)})}]$
can be calculated. Thus the solution of the master equation yields
the diagonal elements of the density matrix in the energy basis as:

\begin{equation}
R(t)=S\, diag(\xi(J_{11}^{(d)},t),\xi(J_{22}^{(d)},t),\xi(J_{33}^{(d)},t),\xi(J_{44}^{(d)},t))\, S^{-1}R(0)=P(t)\, R(0),\nonumber
\end{equation}
where $R(t)=(\rho_{11}(t),\rho_{22}(t),\rho_{33}(t),\rho_{44}(t))^{T}$.
In the energy basis, the Lindbladian corresponding to the non-diagonal elements
in master equation (\ref{master equation}) is in Jordan( diagonal)
form:
\begin{equation}
\mathcal{L}^{nondiag}=J^{(n)}=diag(J_{11}^{(n)},J_{22}^{(n)},J_{33}^{(n)},J_{44}^{(n)})=diag(-2i\xi,2i\xi,-2i\eta,2i\eta)-\half(X_{1}+Y_{2})I,\nonumber
\end{equation}
where $I$ denotes a $4\times4$ identity matrix. Thus the eigenvalues
of Lindbladian of non-diagonal elements is determined as $\lambda_{i}^{(n)}=J_{ii}^{(n)}$
and hence the function $\xi_{i}^{(n)}(t)=\xi(J_{ii}^{(n)},t)=Lap^{-1}[\frac{1}{s-J_{ii}^{(n)}k(s-J_{ii}^{(n)})}]$
can be obtained. The non-diagonal elements of density matrix in the
later time and in the energy basis can be calculated as:
\begin{equation}
Q(t)=diag(\xi(J_{11}^{(n)},t),\xi(J_{22}^{(n)},t),\xi(J_{33}^{(n)},t),\xi(J_{44}^{(n)},t))\, Q(0),\nonumber
\end{equation}
with $Q(t)=(\rho_{12}(t),\rho_{21}(t),\rho_{34}(t),\rho_{43}(t))^{T}$.

Now, the dynamics of reduced density operator of system is determined if the memory function (kernel) is determined. In the following we assume an exponentially damping function for the kernel with the form:
\begin{equation}
k(t)=\gamma_{0}e^{-\gamma_{0}t},\label{kernel}
\end{equation}
where $\gamma_{0}^{-1}$ denotes the characteristic time of the environment's
memory function (also called ``coarse-graining time''). Therefore
we have $\xi(\lambda_{i},t)=\frac{\gamma_{0}e^{\lambda_{i}t}+\lambda_{i}e^{-\gamma_{0}t}}{\lambda_{i}+\gamma_{0}}.$
Consequently, the diagonal terms of density matrix in the energy basis
can be obtained as:
\begin{eqnarray}
\rho_{i\, i}(t)=\sum_{j=1}^{4}p_{i\, j}\,\rho_{j\, j}(0),\label{R t}
\end{eqnarray}
 where $p_{ij}$ are elements of matrix $P(t)=[p_{ij}]_{4\times4}$
and are given in the appendix A, explicitly. The non-diagonal element
of density matrix in the energy basis can be written as:

\begin{eqnarray}\label{nondiagonal}
\,\rho_{1\,2}(t)=\frac{e^{-t\gamma_{0}}\left(X_{1}+Y_{2}-2(e^{-\half t(X_{1}+Y_{2}-2(\gamma_{0}-2i\xi))}\gamma_{0}-2i\xi)\right)}{X_{1}+Y_{2}-2(\gamma_{0}-2i\xi)}\rho_{1\,2}(0),\,\,\,\,\,\,\,\,\rho_{2\,1}(t)=\rho_{1\,2}^{*}(t),\nonumber \\
\rho_{3\,4}(t)=\frac{e^{-t\gamma_{0}}\left(X_{1}+Y_{2}-2(e^{-\half t(X_{1}+Y_{2}-2(\gamma_{0}-2i\eta))}\gamma_{0}-2i\eta)\right)}{X_{1}+Y_{2}-2(\gamma_{0}-2i\eta)}\rho_{3\,4}(0),\,\,\,\,\,\,\,\,\rho_{4\,3}(t)=\rho_{3\,4}^{*}(t).\\
\nonumber 
\end{eqnarray}
The spectrum (\ref{spectrum})becomes degenerate at $\xi=\eta$, for which the above solution is not valid. The state
of the system is not well defined at this critical point. This critical
point assigns a critical value for the parameters of the system such
as critical magnetic field ($B_{c}$), critical parameter of inhomogeneity
of magnetic field ($b_{c}$), critical spin-orbit interaction parameter
($D_{c}$) and etc. Indeed quantum phase transition may be occurs at this critical point and hence the amounts of quantum correlation of the system
changes abruptly when the parameters cross their critical values. The behavior of thermal entanglement at this point is studied in \cite{HamidPRA2008}. For the case of memory-less evolution {\it i.e.} $ k(t)=\delta(t) $ or $ \gamma \rightarrow 0 $ and also for the asymptotic large times {\it i.e.} $t \rightarrow \infty$ the evolution reduce to the Markovian case.
\subsection*{Asymptotic case}
The decoherence induced by environments
does not prevent the creation of a steady state level of quantum correlation,
regardless of the initial state of the system.  Due to Eq. (\ref{kernel}) the effects of memory decrease by time and hence the evolution becomes Markovian, at the large time limit.
At the large time limit, the non-diagonal elements (\ref{nondiagonal})
vanish and $\hat{\rho}(t)$ converges to a diagonal density matrix (in the energy basis):
\begin{eqnarray}
\hat{\rho}^{\infty}=\hat{\rho}_{asymptotic}=\lim_{t\rightarrow\infty}\hat{\rho}(t)=\frac{1}{X_{1}Y_{2}}\,\,\,\textrm{diagonal}(X_{1}^{+}Y_{2}^{+},X_{1}^{-}Y_{2}^{-},X_{1}^{-}Y_{2}^{+},X_{1}^{+}Y_{2}^{-}),\label{rho asym1 }
\end{eqnarray}
which is time independent. Therefore, there is a stationary state which the system tends asymptotically. This asymptotic state is independent on the initial conditions due to forgetful treatment of environment in the Markovian regime. There is an interesting limiting case for which the coupled quantum dots are in contact with the reservoirs at identical
temperatures ($\beta_{1}=\beta_{2}=\beta$). In this case, it is easy
to show that the reduced density matrix $\hat{\rho}^{\infty}$ takes
the thermodynamic canonical form for a system described by the Hamiltonian
$\hat{H}_{S}$ at temperature $T=\beta^{-1}$ \textit{i.e.} $\hat{\rho}^{\infty}(\Delta T=0)\equiv\hat{\rho}_{T}=\frac{e^{-\beta H_{S}}}{Z}$, where $Z=Tr(e^{-\beta H_{S}})$ is the partition function. Thermal entanglement and thermal discord properties of such systems
have been studied substantially in Refs. \cite{HamidPRA2008, WerlangPRA2010}, respectively.

\section{Results and Discussion}
\begin{figure}[ht!]
\centering
\includegraphics[width=17cm]{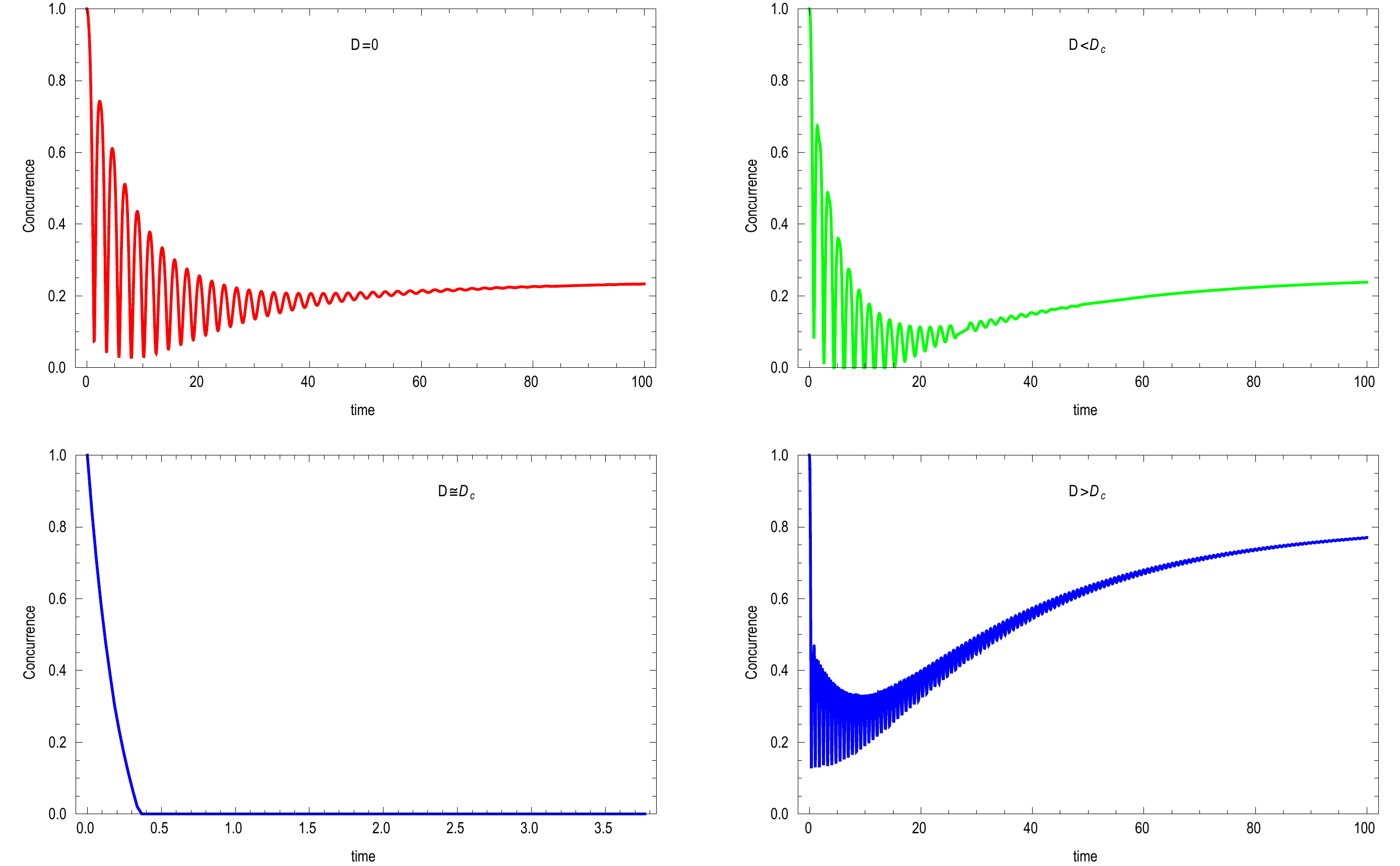}
\caption{(Color online): Dynamical behavior of the concurrence for a maximally entangled initial state for different values of $D$ around $D_c \approx 1.68$. Here $J = 1 $, $\chi = 0.9 $, $B = 2 $, $b = 1 $, $T_ 1 = 1.25 $, $T_ 2 = 0.75 $ and $\frac{\gamma_0} {\bar{\gamma}}= 200 $. All parameters are dimensionless.}
\label{figure1}
\end{figure}

\begin{figure}[ht!]
\centering
\includegraphics[width=17cm]{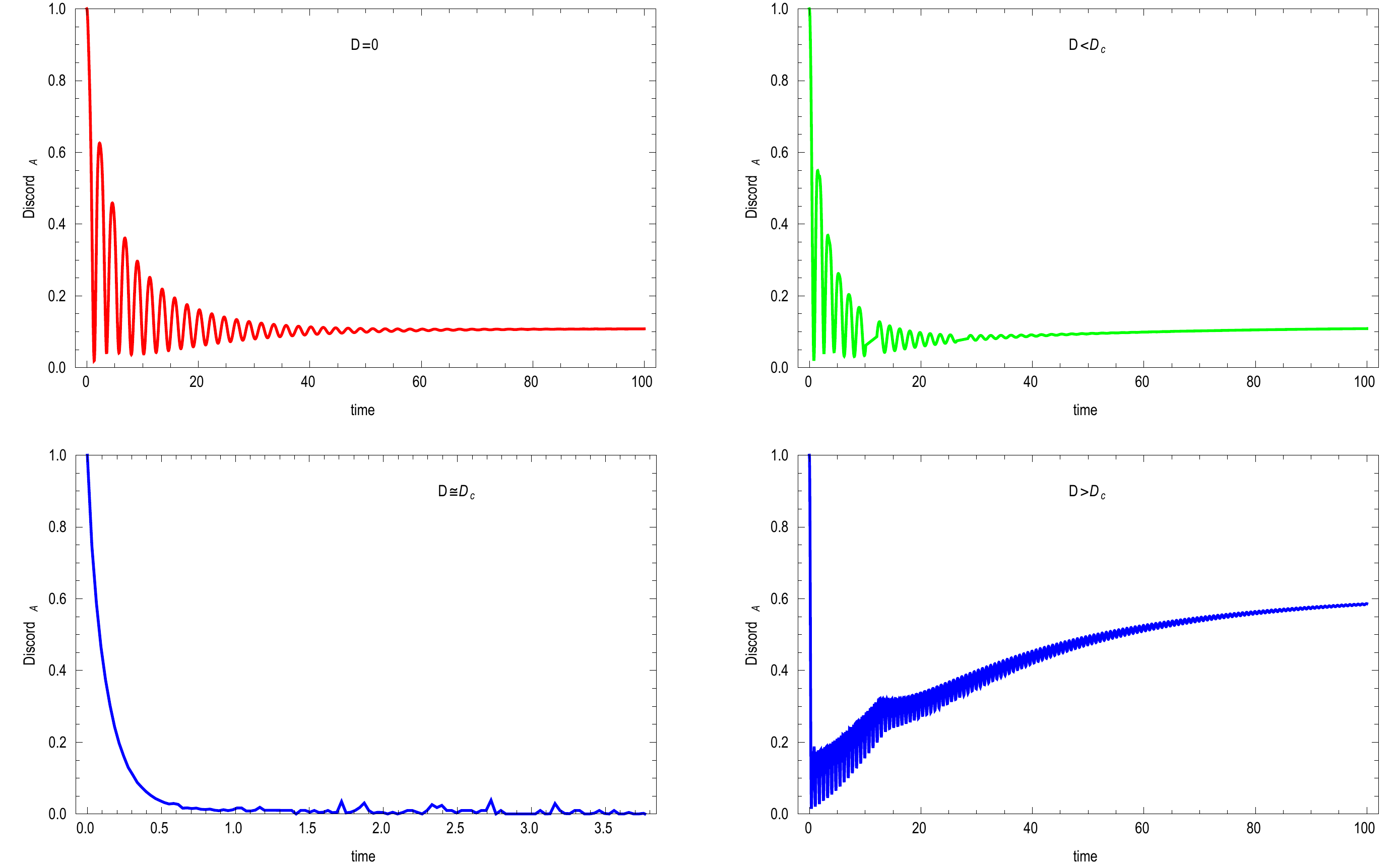}
\caption{(Color online): Dynamical behavior of the left discord for a maximally entangled initial
state for different values of  $D$ around $D_c \approx 1.68 $. Here $J = 1 $, $\chi = 
 0.9 $, $B = 2 $, $b = 1 $, $T_ 1 = 1.25 $, $T_ 2 = 0.75 $ and $\frac{\gamma_0} {\bar{\gamma}}= 200 $. All parameters are dimensionless.}
\label{figure2}
\end{figure}

\begin{figure}[ht!]
\centering
\includegraphics[width=17cm]{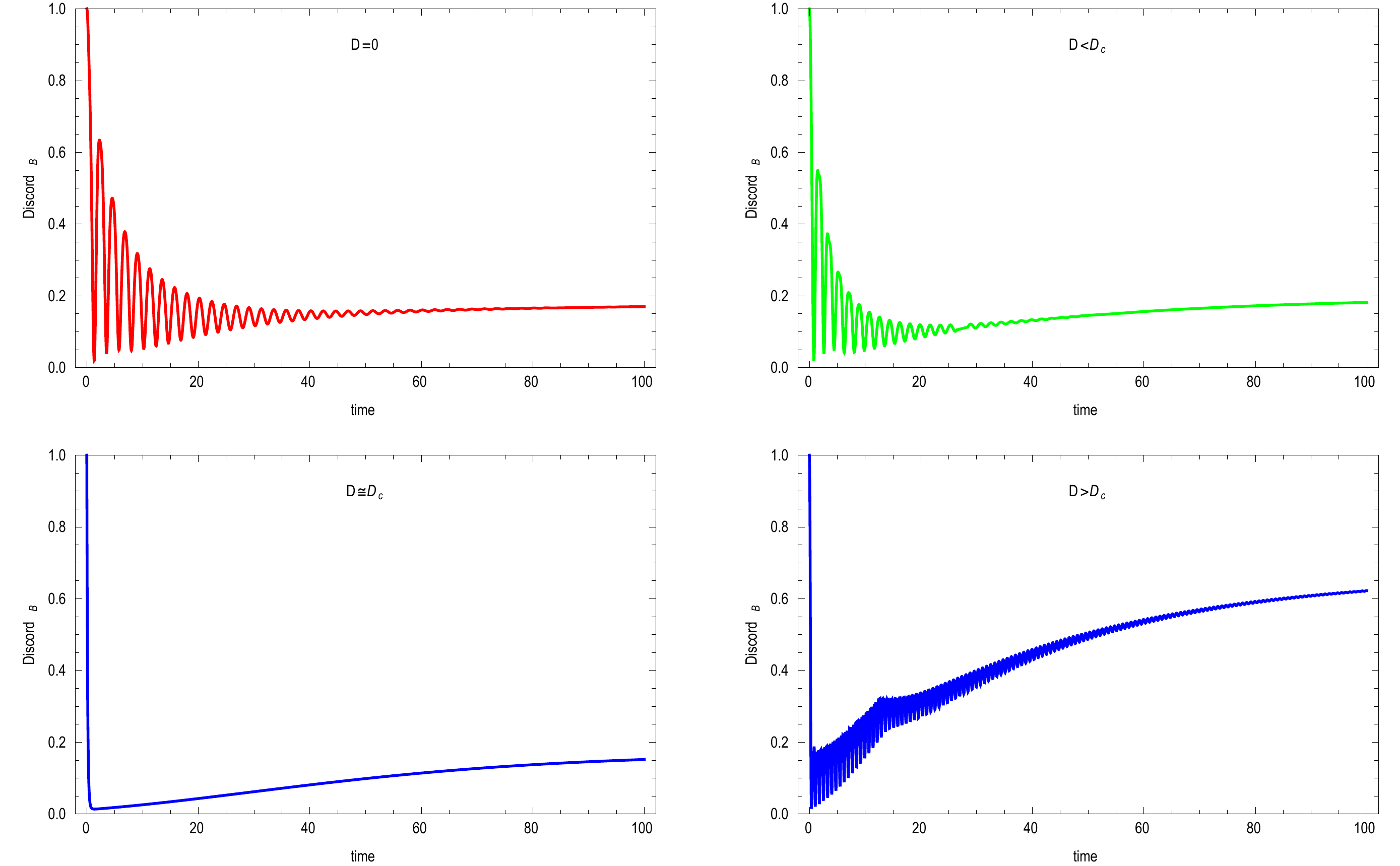}
\caption{(Color online): Dynamical behavior of the right discord for a maximally entangled initial state for different values of $D$ around $D_c \approx 1.68 $. Here $J = 1 $, $\chi = 
 0.9 $, $B = 2 $, $b = 1 $, $T_ 1 = 1.25 $, $T_ 2 = 0.75 $ and $\frac{\gamma_0} {\bar{\gamma}}= 200 $. All parameters are dimensionless. }
\label{figure3}
\end{figure}

\begin{figure}[ht!]
\centering
\includegraphics[width=17cm]{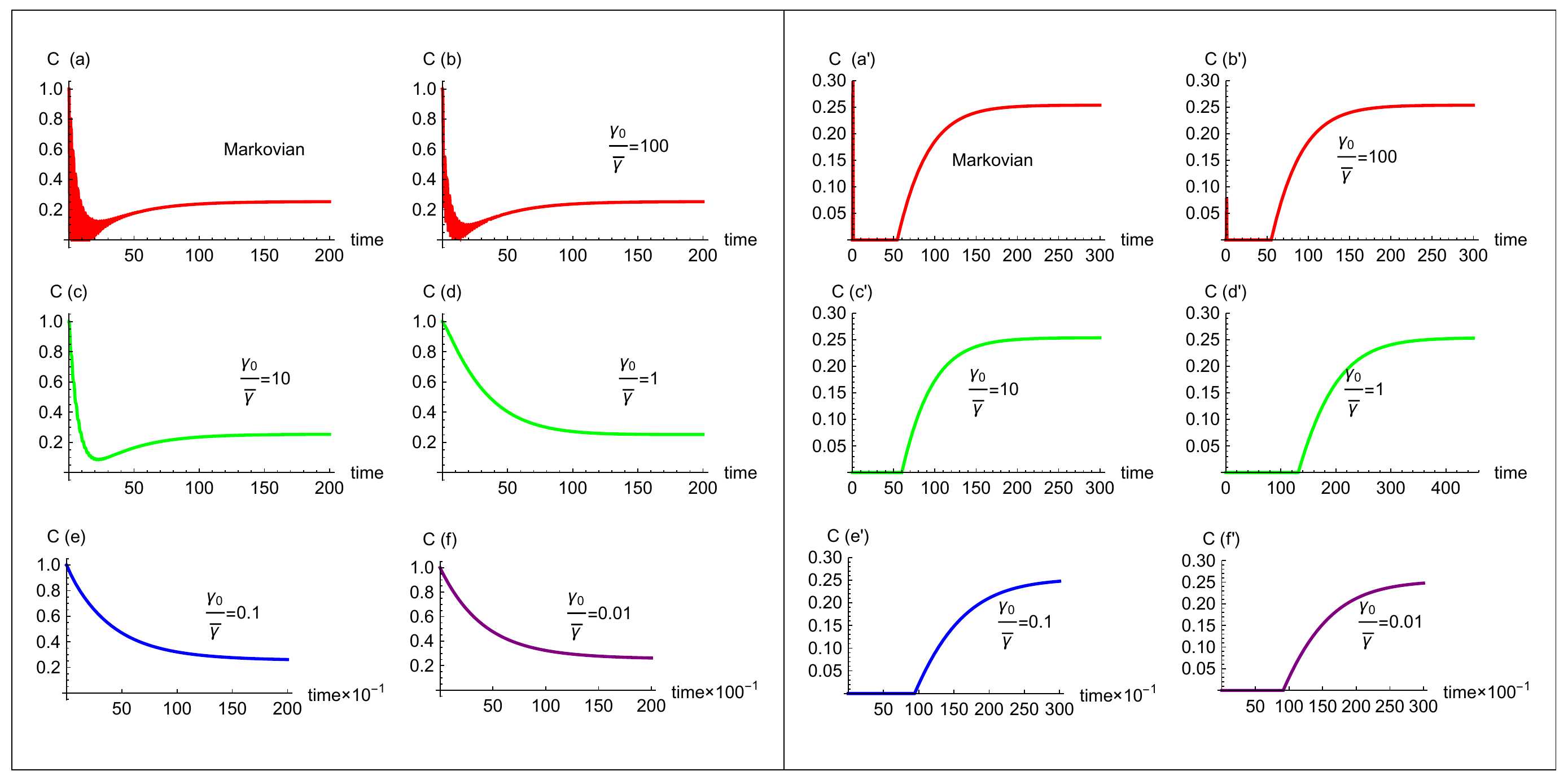}
\caption{(Color online): Dynamics of the concurrence for different values of $\frac{\gamma_0} {\bar{\gamma}}$ when system is initially in a maximally entangled state (left graphs) and a non zero-discord separable state (right graphs). Here $J = 1 $, $\chi = 0.9 $, $B = 2 $, $b = 1 $, $D=1$, $T_ 1 = 1.25 $ and $T_ 2 = 0.75 $. All parameters are dimensionless. }
\label{figure4}
\end{figure}

\begin{figure}[ht!]
\centering
\includegraphics[width=17cm]{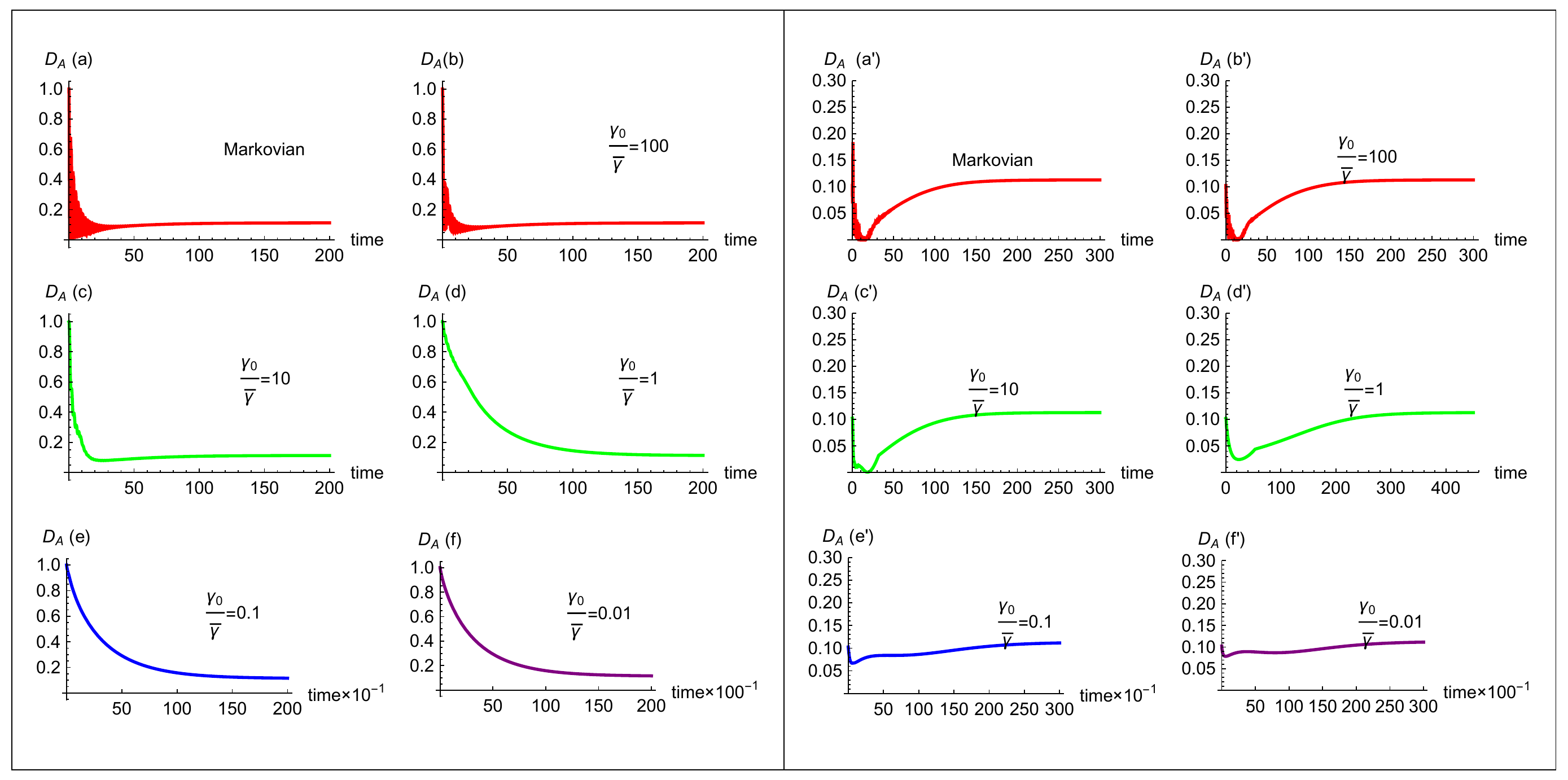}
\caption{(Color online): Dynamics the left discord for different values of $\frac{\gamma_0} {\bar{\gamma}}$ when system is initially in a maximally entangled state (left graphs) and a non zero-discord separable state (right graphs). Here $J = 1 $, $\chi = 0.9 $, $B = 2 $, $b = 1 $, $D=1$, $T_ 1 = 1.25 $ and $T_ 2 = 0.75 $. All parameters are dimensionless. }
\label{figure5}
\end{figure}

\begin{figure}[ht!]
\centering
\includegraphics[width=17cm]{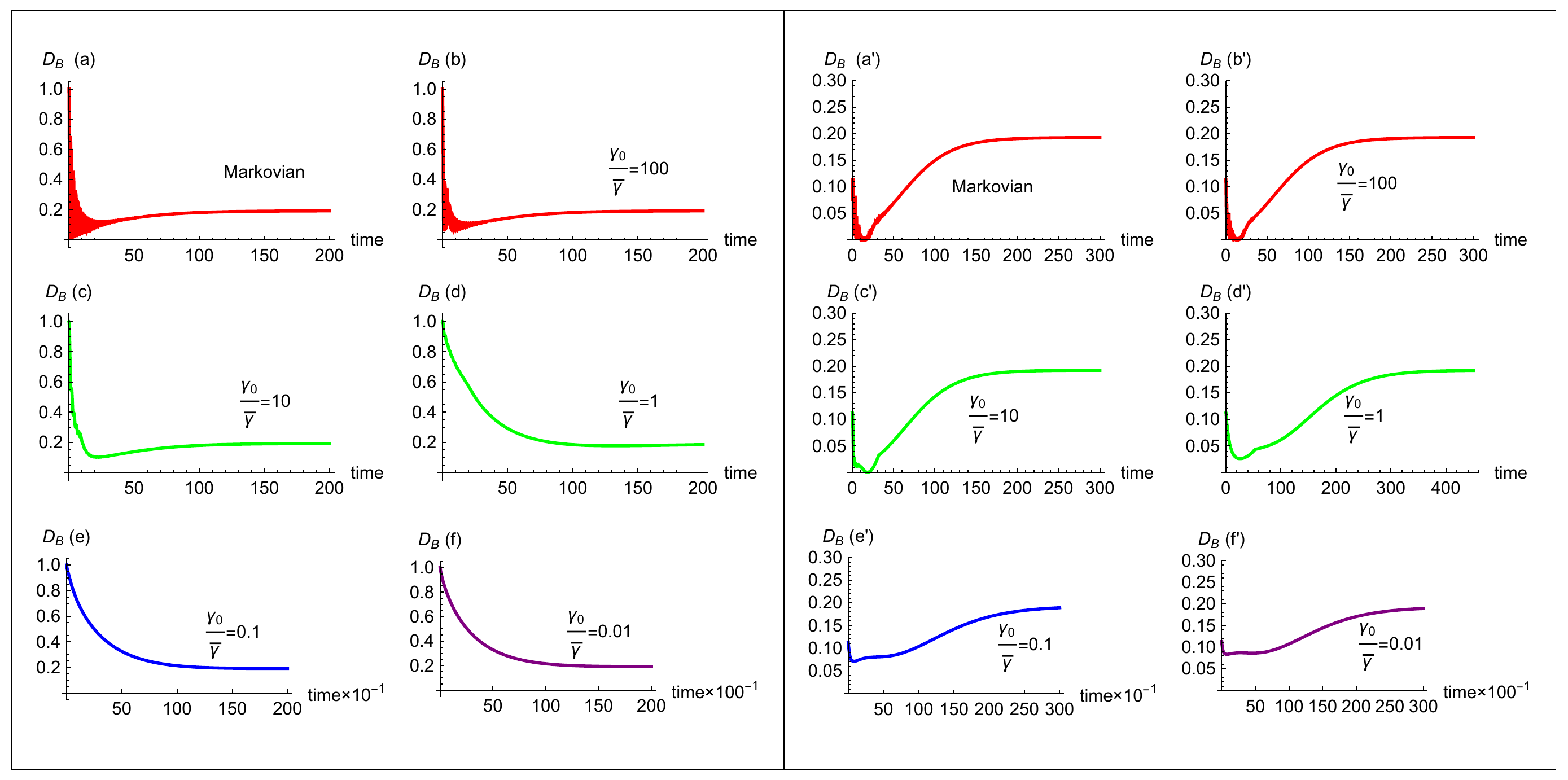}
\caption{(Color online): Dynamics  of the right discord for different values of $\frac{\gamma_0} {\bar{\gamma}}$ when system is initially in a maximally entangled state (left graphs) and a non zero-discord separable state (right graphs). Here $J = 1 $, $\chi = 0.9 $, $B = 2 $, $b = 1 $, $D = 1 $, $T_ 1 = 1.25 $ and $T_ 2 = 0.75 $. All parameters are dimensionless. }
\label{figure6}
\end{figure}

\begin{figure}[ht!]
\centering
\includegraphics[width=17cm]{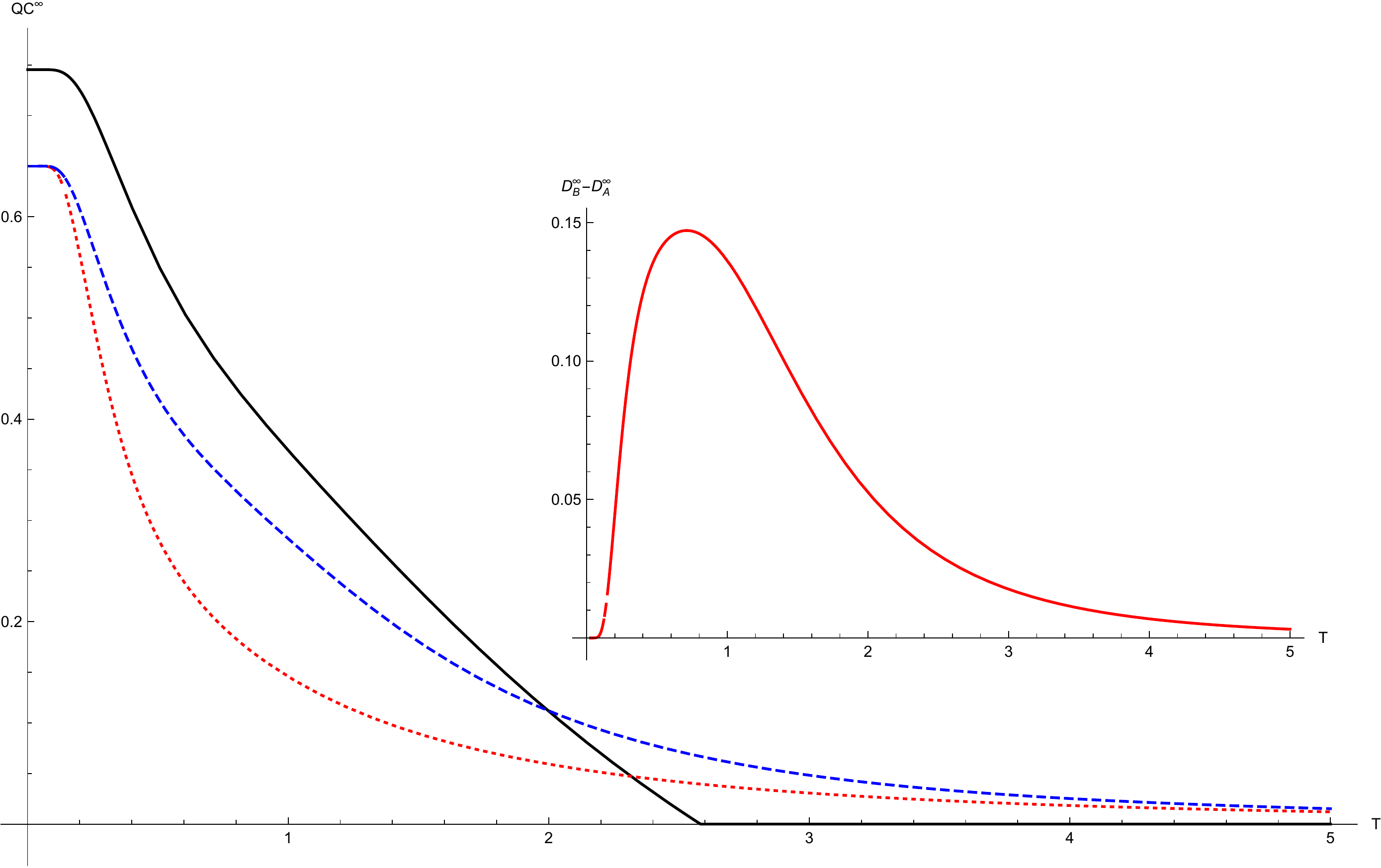}
\caption{(Color online): (Color online) The asymptotic concurrence (black solid line), 
the asymptotic left discord (red dotted line) and the asymptotic \
right discord (blue dashed line) vs. temperature, $T=T_1=T_2$. 
  Inset : $D_B^\infty - D_A^\infty$ vs. $T$. Here $J = 1 $, $\chi = 
 0.9 $, $B = b = 2 $ and  $D=2$. All parameters are dimensionless. }
\label{figure7}
\end{figure}

\begin{figure}[ht!]
\centering
\includegraphics[width=17cm]{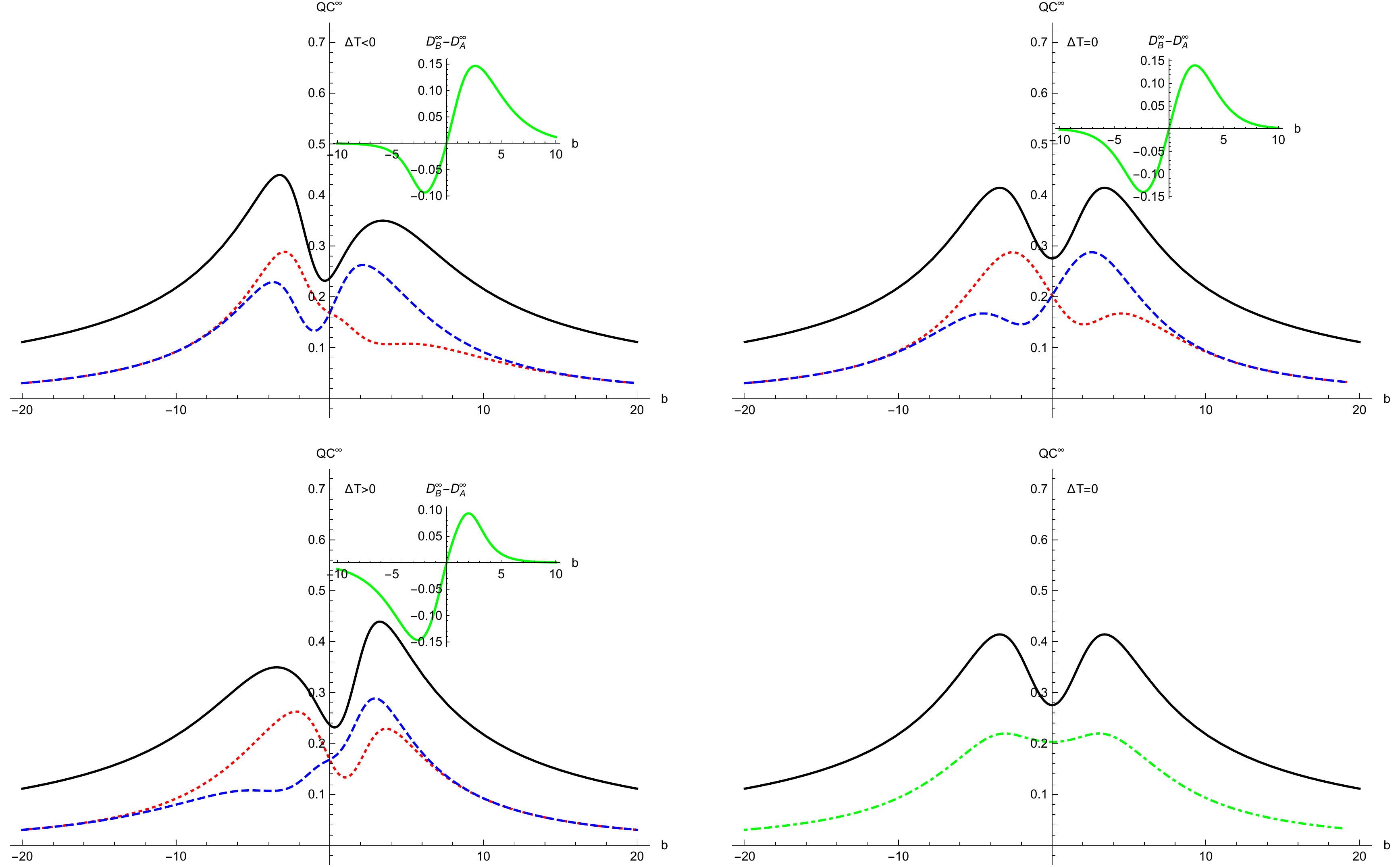}
\caption{(Color online): Influence of geometry of connection on the asymptotic quantum correlations : the asymptotic concurrence (black solid line), the asymptotic left discord (red dotted line) and the asymptotic right discord (blue dashed line). Insets : $D_B^\infty-D_A^\infty$ vs. b for different values of $\Delta T$. The last graph depicts the behavior of the asymptotic 
concurrence (black solid line) and the averaged asymptotic discord ($\frac {D_B^\infty + D_A^\infty} {2} $) (green dot-dashed line) vs. $b$ in the thermal equilibrium condition. Here $J = 1 $, $\chi = 0.9 $, $B = 2 $ and $D = 2 $. All parameters are dimensionless.}
\label{figure8}
\end{figure}

Knowing the density matrix, one can calculate the concurrence, as a measure of entanglement and the quantum discord, as a measure of quantum correlation. Evidently, the results depend on the parameters involved. This prevents one from writing an analytic expression for the concurrence and/or the discord, but it is possible to calculate them for a given set of the parameters. Influence of a parameter on the dynamical and asymptotical behavior of quantum correlations could be studied by drawing their variation versus the mentioned parameter when the other parameters are fixed. The results are depicted in Figures \ref{figure1}-\ref{figure8}.
Figs. \ref{figure1}-\ref{figure6} compare the time evolution of the concurrence and the quantum discord and Figs.\ref{figure7} and \ref{figure8} depict the asymptotic concurrence and quantum discord versus the system and environment parameters. The results of Figs. \ref{figure1}-\ref{figure6} show that all type of considered quantum correlations reach a steady value after some coherent oscillations. These coherent oscillations, are due to competition of the unitary and dissipative terms in master equation (\ref{master equation}). Due to the kernel (\ref{kernel}), the environment losses its memory during the evolution and hence the dynamics tends to the  Markovian case at asymptotically large time limit. 
Figures \ref{figure1}-\ref{figure3} depict influence of spin-orbit parameter, $D$ on the post-Markovian dynamics of the concurrence, left and right discord, respectively for maximally entangled initial state $\ket{\psi^+}=\frac{1}{\sqrt{2}}(\ket{01}+\ket{10})$. The results show that increasing $D$ improves the amount of steady state quantum correlation. Figures \ref{figure4}-\ref{figure6} show the dynamics of concurrence, left and right discord, respectively in different dynamical regimes for maximally entangled and also for non zero-discord separable initial states. The results show that the initial coherent oscillations are indicator of Markovianity of evolution and disappear in non-Markovian regime and also steady state level of quantum correlation achieves at earlier time for Markovian case. Figure \ref{figure7} illustrates the asymptotic quantum correlation vs. temperature when two bathes are held in the same temperature. This figure reveals that the asymptotic entanglement vanishes above a critical temperature (entanglement sudden death). But quantum discord sudden death does not never occurs. This is due to the fact that the set of zero-discord states has no volume in the state space \textit{i.e.} almost all quantum states possess quantum discord \cite{FerraroPRA2010}. Since each qubit experiences different magnetic fields, the symmetry of asymptotic right and left discords breaks for $T\geq 0$. Variation of the asymptotic quantum correlation is depicted in figure \ref{figure8} for differnt ways of connections. Because each qubit is held in its own temperature and experiences different magnetic field, there are two different ways for connecting the quantum dots to their bathes \cite{HamidEPJD2010}: (i) \textquotedbl\textit{direct geometry}{}\textquotedbl{}; where a high temperature bath couples to the quantum dots which is in the stronger magnetic field i.e. $b\Delta T>0$ and (ii) \textquotedbl{}\textit{indirect geometry}\textquotedbl{};
where a high temperature bath couples to the quantum dot which is in the weaker magnetic field i.e. $b\Delta T<0$. The results show that inhomogeneity of magnetic field removes the degeneracy of left and right discord and the amount of this symmetry breaking depends on the temperature difference between bathes. This figure reveals that, if the measurement performed on the qubit which is in the stronger magnetic field, higher amount of asymptotic quantum discord could be achieved. So, the geometry of connection determines the amount of asymptotic quantum correlation an hence is important.

\section{Conclusion}
The Dynamics of non-equilibrium thermal entanglement and thermal discord
of an open two-qubit system is investigated. The inter-qubit interaction
is considered as the Heisenberg interaction in the presence of inhomogeneous
magnetic field and spin-orbit interaction, raised from the Dzyaloshinski-
Moriya (DM) anisotropic anti-symmetric interaction. Each qubit interacts
with a separate thermal reservoir which is held in its own temperature.
For physical realization of the model we address to the spin states
of two electrons which are confined in two coupled quantum dots, respectively.
The dots are assumed to biased via different sources and drains and
hence experience different environments. The effects of the parameters
of the model, including the parameters of the system (especially,
the parameter of the spin-orbit interaction, $D$, and magnetic field inhomogeneity, $b$) and environmental parameters (particularly, mean temperature $T_{M}$ and temperature difference
$\Delta T$), on the dynamics of the system is investigated, by solving
the quantum Markov-Born master equation of the system. Tracing the
dynamics of the system allowed us to distinguish between the quantum
correlation produced by the inter-qubit interaction and/or by the environment. Decoherence induced by thermal bathes are competing with inter-qubit
interaction terms leading to the system evolves to an asymptotic steady
state. The size of the entanglement and discord of this steady state
and also the dynamical behavior of the entanglement depend on the
parameters of the model and also on the geometry of connection. The results reveal that increasing the size of DM interaction, $D$, enhances the amount of all asymptotic quantum correlation measures.
Also the results show that
the asymptotic entanglement of the system dies above a critical temperature $T_{cr.}$ and entanglement sudden death occurs. The size of $T_{cr.}$ and the amount of asymptotic entanglement
can be enhanced by choosing a suitable value of $D$ and the temperature difference $\Delta T$.  On the other hand  the results show that thermal discord could live in higher temperatures than thermal entanglement and quantum discord sudden death does not occurs. Also, introducing magnetic field inhomogeneity breaks the symmetry between left and right discord. The results show that if the magnetic field applied on right(left) qubit is greater then the size of right(left) discord is higher. Furthermore, we find that choosing proper geometry of connection is important for creating and maintaining the quantum correlation.

\begin{acknowledgments}
The author wish to thank The Office of Graduate Studies and
Research Vice President of The University of Isfahan for their support. 
\end{acknowledgments}

\newpage
\appendix
\section{elements of marix P}
The elements of matrix $P(t)=[p_{ij}]_{4\times4}$ in the equation
(\ref{R t}) can be written as follow:
\begin{eqnarray}
p_{1\,1} & = & \frac{1}{X_{1}Y_{2}}[\frac{X_{1}^{-}Y_{2}^{+}(e^{-t\gamma_{0}}X_{1}-e^{-tX_{1}}\gamma_{0})}{X_{1}-\gamma_{0}}+\frac{X_{1}^{+}Y_{2}^{-}\left(e^{-t\gamma_{0}}Y_{2}-e^{-tY_{2}}\gamma_{0}\right)}{Y_{2}-\gamma_{0}}\nonumber\\
 & + & \frac{e^{-t\gamma_{0}}X_{1}^{-}Y_{2}^{-}\left(X_{1}+Y_{2}-e^{-t\left(X_{1}+Y_{2}-\gamma_{0}\right)}\gamma_{0}\right)}{X_{1}+Y_{2}-\gamma_{0}}+X_{1}^{+}Y_{2}^{+}],\nonumber\\
p_{1\,2} & = & \frac{X_{1}^{+}Y_{2}^{+}}{X_{1}Y_{2}}[\frac{e^{-t\gamma_{0}}X_{1}Y_{2}(X_{1}+Y_{2}-2\gamma_{0})}{(X_{1}-\gamma_{0})(X_{1}+Y_{2}-\gamma_{0})(\gamma_{0}-Y_{2})}+\frac{e^{-tX_{1}}\gamma_{0}}{X_{1}-\gamma_{0}}+\frac{e^{-tY_{2}}\gamma_{0}}{Y_{2}-\gamma_{0}}\nonumber\\
 & - & \frac{e^{-t(X_{1}+Y_{2})}\gamma_{0}}{X_{1}+Y_{2}-\gamma_{0}}+1]\nonumber\\
p_{1\,3} & = & \frac{X_{1}^{+}}{X_{1}Y_{2}}[-\frac{\left(e^{-tX_{1}}\gamma_{0}-e^{-t\gamma_{0}}X_{1}\right)Y_{2}^{+}}{\gamma_{0}-X_{1}}+\frac{Y_{2}^{-}\left(e^{-tY_{2}}\gamma_{0}-e^{-t\gamma_{0}}Y_{2}\right)}{\gamma_{0}-Y_{2}}\nonumber\\
 & + & \frac{e^{-t\gamma_{0}}Y_{2}^{-}\left(X_{1}+Y_{2}-e^{-t(X_{1}+Y_{2}-\gamma_{0})}\gamma_{0}\right)}{-X_{1}-Y_{2}+\gamma_{0}}+Y_{2}^{+}],\nonumber\\
p_{1\,4} & = & \frac{Y_{2}^{+}}{X_{1}Y_{2}}[-\frac{\left(e^{-tY_{2}}\gamma_{0}-e^{-t\gamma_{0}}Y_{2}\right)X_{1}^{+}}{\gamma_{0}-Y_{2}}+\frac{X_{1}^{-}\left(e^{-tX_{1}}\gamma_{0}-e^{-t\gamma_{0}}X_{1}\right)}{\gamma_{0}-X_{1}}\nonumber\\
 & + & \frac{e^{-t\gamma_{0}}X_{1}^{-}\left(X_{1}+Y_{2}-e^{-t(X_{1}+Y_{2}-\gamma_{0})}\gamma_{0}\right)}{-X_{1}-Y_{2}+\gamma_{0}}+X_{1}^{+}],\nonumber\\
p_{2\,1} & = & \frac{X_{1}^{-}Y_{1}^{-}}{X_{1}Y_{2}}[\frac{e^{-t\gamma_{0}}X_{1}Y_{2}(X_{1}+Y_{2}-2\gamma_{0})}{(X_{1}-\gamma_{0})(X_{1}+Y_{2}-\gamma_{0})(\gamma_{0}-Y_{2})}+\frac{e^{-tX_{1}}\gamma_{0}}{X_{1}-\gamma_{0}}+\frac{e^{-tY_{2}}\gamma_{0}}{Y_{2}-\gamma_{0}}\nonumber\\
 & - & \frac{e^{-t(X_{1}+Y_{2})}\gamma_{0}}{X_{1}+Y_{2}-\gamma_{0}}+1],\nonumber\\
p_{2\,2} & = & \frac{1}{X_{1}Y_{2}}[\frac{X_{1}^{+}Y_{2}^{-}(e^{-t\gamma_{0}}X_{1}-e^{-tX_{1}}\gamma_{0})}{X_{1}-\gamma_{0}}+\frac{X_{1}^{-}Y_{2}^{+}\left(e^{-t\gamma_{0}}Y_{2}-e^{-tY_{2}}\gamma_{0}\right)}{Y_{2}-\gamma_{0}}\nonumber\\
 & + & \frac{e^{-t\gamma_{0}}X_{1}^{+}Y_{2}^{+}\left(X_{1}+Y_{2}-e^{-t\left(X_{1}+Y_{2}-\gamma_{0}\right)}\gamma_{0}\right)}{X_{1}+Y_{2}-\gamma_{0}}+X_{1}^{-}Y_{2}^{-}],\nonumber\\
p_{2\,3} & = & \frac{Y_{2}^{-}}{X_{1}Y_{2}}[\frac{X_{1}^{+}\left(e^{-tX_{1}}\gamma_{0}-e^{-t\gamma_{0}}X_{1}\right)}{\gamma_{0}-X_{1}}-\frac{\left(e^{-tY_{2}}\gamma_{0}-e^{-t\gamma_{0}}Y_{2}\right)X_{1}^{-}}{\gamma_{0}-Y_{2}}\nonumber\\
 & + & \frac{e^{-t\gamma_{0}}X_{1}^{+}\left(X_{1}+Y_{2}-e^{-t(X_{1}+Y_{2}-\gamma_{0})}\gamma_{0}\right)}{-X_{1}-Y_{2}+\gamma_{0}}+X_{1}^{-}],\nonumber\\
p_{2\,4} & = & \frac{X_{1}^{-}}{X_{1}Y_{2}}[-\frac{\left(e^{-tX_{1}}\gamma_{0}-e^{-t\gamma_{0}}X_{1}\right)Y_{2}^{-}}{\gamma_{0}-X_{1}}+\frac{Y_{2}^{+}\left(e^{-tY_{2}}\gamma_{0}e^{-t\gamma_{0}}Y_{2}\right)}{\gamma_{0}-Y_{2}}\nonumber\\
 & + & \frac{e^{-t\gamma_{0}}Y_{2}^{+}\left(X_{1}+Y_{2}-e^{-t(X_{1}+Y_{2}-\gamma_{0})}\gamma_{0}\right)}{-X_{1}-Y_{2}+\gamma_{0}}+Y_{2}^{-}],\nonumber
\end{eqnarray}

\begin{eqnarray}
p_{3\,1} & = & \frac{X_{1}^{-}}{X_{1}Y_{2}}[-\frac{\left(e^{-tX_{1}}\gamma_{0}-e^{-t\gamma_{0}}X_{1}\right)Y_{2}^{+}}{\gamma_{0}-X_{1}}+\frac{Y_{2}^{-}\left(e^{-tY_{2}}\gamma_{0}-e^{-t\gamma_{0}}Y_{2}\right)}{\gamma_{0}-Y_{2}}\nonumber \\
 & + & \frac{e^{-t\gamma_{0}}Y_{2}^{-}\left(X_{1}+Y_{2}-e^{-t(X_{1}+Y_{2}-\gamma_{0})}\gamma_{0}\right)}{-X_{1}-Y_{2}+\gamma_{0}}+Y_{2}^{+}],\nonumber \\
p_{3\,2} & = & \frac{Y_{2}^{+}}{X_{1}Y_{2}}[\frac{X_{1}^{+}\left(e^{-tX_{1}}\gamma_{0}-e^{-t\gamma_{0}}X_{1}\right)}{\gamma_{0}-X_{1}}-\frac{\left(e^{-tY_{2}}\gamma_{0}-e^{-t\gamma_{0}}Y_{2}\right)X_{1}^{-}}{\gamma_{0}-Y_{2}}\nonumber \\
 & + & \frac{e^{-t\gamma_{0}}X_{1}^{+}\left(X_{1}+Y_{2}-e^{-t(X_{1}+Y_{2}-\gamma_{0})}\gamma_{0}\right)}{-X_{1}-Y_{2}+\gamma_{0}}+X_{1}^{-}],\nonumber \\
p_{3\,3} & = & \frac{1}{X_{1}Y_{2}}[\frac{X_{1}^{+}Y_{2}^{+}(e^{-t\gamma_{0}}X_{1}-e^{-tX_{1}}\gamma_{0})}{X_{1}-\gamma_{0}}+\frac{X_{1}^{-}Y_{2}^{-}\left(e^{-t\gamma_{0}}Y_{2}-e^{-tY_{2}}\gamma_{0}\right)}{Y_{2}-\gamma_{0}}\nonumber \\
 & + & \frac{e^{-t\gamma_{0}}X_{1}^{+}Y_{2}^{-}\left(X_{1}+Y_{2}-e^{-t\left(X_{1}+Y_{2}-\gamma_{0}\right)}\gamma_{0}\right)}{X_{1}+Y_{2}-\gamma_{0}}+X_{1}^{-}Y_{2}^{+}],\nonumber \\
p_{3\,4} & = & \frac{X_{1}^{-}Y_{2}^{+}}{X_{1}Y_{2}}[\frac{e^{-t\gamma_{0}}X_{1}Y_{2}(X_{1}+Y_{2}-2\gamma_{0})}{(X_{1}-\gamma_{0})(X_{1}+Y_{2}-\gamma_{0})(\gamma_{0}-Y_{2})}+\frac{e^{-tX_{1}}\gamma_{0}}{X_{1}-\gamma_{0}}+\frac{e^{-tY_{2}}\gamma_{0}}{Y_{2}-\gamma_{0}}\nonumber \\
 & - & \frac{e^{-t(X_{1}+Y_{2})}\gamma_{0}}{X_{1}+Y_{2}-\gamma_{0}}+1]\nonumber \\
p_{4\,1} & = & \frac{Y_{2}^{-}}{X_{1}Y_{2}}[-\frac{\left(e^{-tY_{2}}\gamma_{0}-e^{-t\gamma_{0}}Y_{2}\right)X_{1}^{+}}{\gamma_{0}-Y_{2}}+\frac{X_{1}^{-}\left(e^{-tX_{1}}\gamma_{0}-e^{-t\gamma_{0}}X_{1}\right)}{\gamma_{0}-X_{1}}\nonumber \\
 & + & \frac{e^{-t\gamma_{0}}X_{1}^{-}\left(X_{1}+Y_{2}-e^{-t(X_{1}+Y_{2}-\gamma_{0})}\gamma_{0}\right)}{-X_{1}-Y_{2}+\gamma_{0}}+X_{1}^{+}],\nonumber \\
p_{4\,2} & = & \frac{X_{1}^{+}}{X_{1}Y_{2}}[-\frac{\left(e^{-tX_{1}}\gamma_{0}-e^{-t\gamma_{0}}X_{1}\right)Y_{2}^{-}}{\gamma_{0}-X_{1}}+\frac{Y_{2}^{+}\left(e^{-tY_{2}}\gamma_{0}e^{-t\gamma_{0}}Y_{2}\right)}{\gamma_{0}-Y_{2}}\nonumber \\
 & + & \frac{e^{-t\gamma_{0}}Y_{2}^{+}\left(X_{1}+Y_{2}-e^{-t(X_{1}+Y_{2}-\gamma_{0})}\gamma_{0}\right)}{-X_{1}-Y_{2}+\gamma_{0}}+Y_{2}^{-}],\nonumber \\
p_{4\,3} & = & \frac{X_{1}^{+}Y_{2}^{-}}{X_{1}Y_{2}}[\frac{e^{-t\gamma_{0}}X_{1}Y_{2}(X_{1}+Y_{2}-2\gamma_{0})}{(X_{1}-\gamma_{0})(X_{1}+Y_{2}-\gamma_{0})(\gamma_{0}-Y_{2})}+\frac{e^{-tX_{1}}\gamma_{0}}{X_{1}-\gamma_{0}}+\frac{e^{-tY_{2}}\gamma_{0}}{Y_{2}-\gamma_{0}}\nonumber \\
 & - & \frac{e^{-t(X_{1}+Y_{2})}\gamma_{0}}{X_{1}+Y_{2}-\gamma_{0}}+1]\nonumber \\
p_{4\,4} & = & \frac{1}{X_{1}Y_{2}}[\frac{X_{1}^{-}Y_{2}^{-}(e^{-t\gamma_{0}}X_{1}-e^{-tX_{1}}\gamma_{0})}{X_{1}-\gamma_{0}}+\frac{X_{1}^{+}Y_{2}^{+}\left(e^{-t\gamma_{0}}Y_{2}-e^{-tY_{2}}\gamma_{0}\right)}{Y_{2}-\gamma_{0}}\nonumber \\
 & + & \frac{e^{-t\gamma_{0}}X_{1}^{-}Y_{2}^{+}\left(X_{1}+Y_{2}-e^{-t\left(X_{1}+Y_{2}-\gamma_{0}\right)}\gamma_{0}\right)}{X_{1}+Y_{2}-\gamma_{0}}+X_{1}^{+}Y_{2}^{-}].\label{p i j}
\end{eqnarray}
Here we have defined $X_{\mu}=X_{\mu}^{+}+X_{\mu}^{-}$ and $Y_{\mu}=Y_{\mu}^{+}+Y_{\mu}^{-}$.
\newpage{}
%%%%%%%%%%%%%%%%%%%%%%%%%%%%%%%%%%%%% Figures %%%%%%%%%%%%%%%%%%%%%%%%%%%%%%%%%%%%%%%%%

%%%%%%%%%%%%%%%%%%%%%%%%%%%%%%%% end of figures %%%%%%%%%%%%%%%%%%%%%%%%%%%%%%%%%%%%%%%%%%%%%%%%%
\end{document}